\documentclass[12pt]{article}
\usepackage{amsmath,amssymb,array,enumitem,fullpage,natbib,svn,url}
\usepackage[usenames]{color}
\usepackage[dvips]{graphicx}
\usepackage[active]{srcltx}
\graphicspath{{./}{eps/}}

\newcommand{\dy}{\thinspace{\text{d}}} 

\newcommand{\bbN}{{\mathbb{N}}}

\newcommand{\cL}{{\mathcal{L}}}

\renewcommand{\P}{{\mathsf{P}}} \newcommand{\E}{{\mathsf{E}}}
 
\newcommand{\V}{{\mathsf{V}}}
\newcommand{\Bi}{\mathsf{Bi}}

\newcommand{\Ex}{\mathsf{Ex}}\newcommand{\Ga}{\mathsf{Ga}}

\newcommand{\NB}{\mathsf{NB}}\newcommand{\No}{\mathsf{No}}

\newcommand{\Po}{\mathsf{Po}}\newcommand{\Un}{\mathsf{Un}}

\newcommand{\hide}[1]{}

\newcommand{\iid}{\mathrel{\mathop{\sim}\limits^{\mathrm{iid}}}}

\newbox\asbox
\setbox\asbox=\hbox{\vrule height 15pt depth3.5pt width0pt}
\def\astrut{\relax\ifmmode\copy\strutbox\else\unhcopy\strutbox\fi}
\newdimen\bsigdep
\def\bSig{{\setbox0\hbox{$\Sigma$}\bsigdep=1\dp0\advance\bsigdep by
 .2\ht0 \rlap{\kern.3\wd0\vrule height1.2\ht0 depth1\bsigdep}\box0}}%
\newcommand{\ans}[1]{\hbox to #1truecm{
    \vrule height 20pt depth3.5pt width0pt
    \leaders\hrule height-2.5pt depth3pt\hfill}}

\newcommand{\ie}{\emph{i.e.{}}}

\newcommand{\half}{{\mathchoice{{\textstyle\frac12}} {{\textstyle\frac12}}
                {{\scriptscriptstyle\frac12}}{{\scriptscriptstyle\frac12}}}}

\newcommand{\Sec}[1]{Section\thinspace(\ref{#1})}

\newcommand{\Eqn}[1]{Eqn\thinspace(\ref{#1})}

\newcommand{\Eqns}[2]{Eqns\thinspace(\ref{#1},\thinspace\ref{#2})}

\newcommand{\Fig}[1]{Figure\thinspace(\ref{#1})}

\renewcommand{\ij}{_{ij}}

\newcommand{\bet}[1]{\left [#1\right ]} 
\newcommand{\set}[1]{\left\{#1\right\}} 
\newcount\ola \newcount\olb \newcount\olc \newcount\old \newcount\ole
\newcount\och\newcount\level

\ifx\newcolumntype\undefined
\else
  \newcolumntype{C}{>{$}c<{$}}
  \newcolumntype{L}{>{$}l<{$}}
  \newcolumntype{R}{>{$}r<{$}}
\fi
\def\OL#1{\par\noindent\hangindent=#1\parindent 
  \kern1\hangindent\ignorespaces}%
\def\ol#1{%
    \level=#1
    \ifcase\level
    \ola=0 \olb=0 \olc=0 \old=0 \ole=0\or         
    \olb=0 \olc=0 \old=0 \ole=0 \advance\ola by 1 
    \gdef\olev{\uppercase\expandafter{\romannumeral\ola}} \or
    \olc=0 \old=0 \ole=0 \advance\olb by 1        
    \och=64 \advance\och by\olb
    \gdef\olev{\char\och}\or
    \old=0 \ole=0 \advance\olc by 1               
    \och=48 \advance\och by\olc
    \gdef\olev{\char\och}\or
    \ole=0 \advance\old by 1                      
    \och=96 \advance\och by\old
    \gdef\olev{\char\och}\or
    \advance\ole by 1                             
    \gdef\olev{\romannumeral\ole} \or
    \message{Outline depth too deep: #1}\fi
    \ifnum\level>0 \OL\level\llap{\olev.\enspace}\ignorespaces\fi}%
\long\def\comment#1/*#2*/{\endcomment}%
\def\endcomment{\relax}%
\newcounter{probno}\newcounter{partno}[probno]
\newcommand{\newpart}[1][0]{\ifnum\value{partno}=0\medskip\else\vfill\fi
  \par\stepcounter{partno}\alph{partno})
  \ifnum#1<0(XC)\fi\ifnum#1=0\quad\fi\ifnum#1>0(#1)~\fi}

\makeatletter 
\multiply\count2 60 \advance\count2 \time
\edef\now{\two@digits{\the\count1}:\two@digits{\the\count2}}
\makeatother

\def\wbox#1#2#3{{\vcenter{\vbox{\hrule height.#3pt
    \hbox{\vrule width.#3pt height#1pt \kern#2pt \vrule width.#3pt}%
                            \hrule height.#3pt}}}}%
\def\Proof.{\medbreak\noindent{\bf Proof.\enspace}}

\ifx\url\undefined
\else
\makeatletter
\def\url@rlwstyle{%
  \@ifundefined{selectfont}{\def\UrlFont{\sf}}
    {\def\UrlFont{\small\ttfamily}}}
\makeatother
\fi
\newif\ifdraft
\newif\ifsame
\newcommand{\strcfstr}[2]{%
  \samefalse
  \begingroup
    \def\1{#1}\def\2{#2}%
    \ifx\1\2\endgroup \sametrue
    \else \endgroup
    \fi}
\def\rev$Revi#1: #2 ${#2}
\def\dat$Dat#1: #2 #3 ${#2}
\def\need#1{\vskip0pt plus#1in\penalty-250\vskip0pt plus-#1in}%
\def\SVNtz$#1 -0#200#3${\global\def\tz{\ifcase#2 GMT\or-1\or-2\or ADT\or
    EDT\or EST \or CST\or MST\or PST\or AKST\or -10\or HST\else -#2\fi}}
{\Mst}{\ensuremath{\hat M^\star}}     
\newcommand{\Qst}{Q^\star}
\newcommand{\Rsti}{\Rst_i}
\newcommand{\Rst}{R^\star}
\newcommand{\Secs}[2]{Sections\thinspace(\ref{#1},\thinspace\ref{#2})}
\newcommand{\Tim}[1]{T_{i{-}#1}}
\newcommand{\Tst}{T^\star}
\newcommand{\bF}{{\bar F}}
\newcommand{\bleed}{{\mathfrak B}} 
\newcommand{\hi}[1]{\mathtt{hi(#1)}}
\newcommand{\imo}{_{i{-}1}}
\newcommand{\kpo}{(k{+}1)}
\newcommand{\lo}[1]{\mathtt{lo(#1)}}
\newcommand{\oo}{_{\text{\tiny old}}}
\newcommand{\on}{_{\text{\tiny new}}}
\newcommand{\oz}{\Rst-\Tst_0}
\newcommand{\malp}{\xi}
\newcommand{\mbet}{\lambda}
\newcommand{\pbeta}[1]{\mathtt{pbeta(#1)}}
\newcommand{\pbinom}[1]{\mathtt{pbinom(#1)}}
\newcommand{\qnbinom}[1]{\mathtt{qnbinom(#1)}}
\renewcommand{\dy}{\textsf{\,day}}
\newcommand{\red}[1]{\textcolor{red}{#1}}
\SVN $Rev: 128 $
\SVNdate $Date: 2015-05-20 06:47:48 -0400 (Wed, 20 May 2015) $
\SVNtz   $Date: 2015-05-20 06:47:48 -0400 (Wed, 20 May 2015) $
\defcitealias{CEC:2007}{CEC, 2007}
\begin{document}
\title{ACME: A Partially Periodic Estimator of\\ Avian \& Chiropteran
  Mortality at Wind Turbines} \author{Robert L. Wolpert\thanks {Robert
    L. Wolpert (\texttt{wolpert@stat.duke.edu, +1-919-684-3275}) is
    Professor of Statistical Science and Professor of Environmental Science
    and Policy, Nicholas School of the Environment at Duke University,
    Durham NC, USA.}}
\date{2015 July 02}
\maketitle

\section{Introduction}\label{s:Intro}
While wind energy has been employed for electricity production since the
1880s, it wasn't until the oil crisis of the 1970s that commercial wind
energy production was pursued actively in the United States.  Wind energy
use has grown rapidly since it began to be promoted as an alternative to
fossil fuels and was accorded sponsorship by the state of California in the
1980s and by the Federal Government beginning in the late 1990s.  Concerns
about avian and chiropteran deaths caused by wind turbines emerged in the
early 1990s \citep {Howe:DiDo:1991}, with widely varying estimates of the
fatality rates, and studies were mounted to assess these rates as early as
1998 \citep {Smal:Thel:2005}.  Aggregate U.S.\ mortality estimates have
been reported ranging from 20,000 to 573,000 birds annually\ \citep
{Eric:John:Etal:2001, Eric:John:Youn:Etal:2005, Loss:Will:Marr:2013,
  Manv:2009, Smal:2013a, Sova:2012}.  High profile lawsuits in such places
as Altamont, CA (2007), Ventura, CA (2012), Nantucket Sound, MA (2012),
Port Clinton, OH (2014) have brought the issue to national prominence.

The na\"\i ve approach to estimating turbine-related avian and chiropteran
mortality--- surveying periodically for bird and bat carcasses in
designated areas near turbines at prescribed time intervals, and scaling
the counts by time interval and study area--- leads to grossly distorted
estimates, for a variety of reasons.  Some carcasses will be removed by
scavengers before the survey, for example; some carcasses may be present
but undetected at the time of the survey; some fatally injured birds or
bats may survive long enough to alight outside the study area; and
carcasses may be discovered whose death arose from other causes or during
other time periods.

A number of investigators have developed modeling approaches leading to
proposed adjustment formulas intended to overcome the distortions and
biases of the na\"\i ve approach \citep{Eric:Stric:etal:1998,
  John:Eric:etal:2003, Shoe:2004, Poll:2007, Huso:2011}, each embodying
slightly different assumptions about the processes affecting carcass
discovery.  The wide variability of these estimation formulas leaves
practitioners uncertain which of them (if any) to use.  Here we explain the
assumptions that underlie four commonly used estimation formulas,
illustrate when each is appropriate and how they differ, and propose a new
model-based Avian and Chiropteran Mortality Estimator called ``ACME'' that
extends all four of them and introduces three new features to improve the
reliability of mortality estimates: the diminishment of Field Technician
(FT) discovery proficiency as carcasses age; the reduced rate of scavenger
removal as carcasses age; and the possibility that some but not all
carcasses present but undiscovered by FTs in one search may be discovered
in a later search.

\section{The Model Underlying the New Estimator}\label{s:model}

Suppose that carcasses arrive in a Poisson stream with intensity $m(t)$
that varies slowly with time $t$ and that they are removed (principally by
scavengers) independently after random times $\tau_j$ with complimentary
CDF $\bF(t)=\P[\tau_j>t]$.  Suppose too that field technicians (FTs) mount
blinded searches at a sequence of times $T_i$ at similar intervals
$I_i=[T_i-\Tim1]$, and that the probability that a carcass of age $\tau$
will be discovered by an FT in such a search is $S(\tau)$ (which may depend
on the carcass age $\tau$, but we are assuming for now that discovery is
statistically independent of the scavenging removal process).  Let $C_i$
denote the (random) number of carcasses actually discovered in the search
at time $T_i$.  Then the expected number of carcasses that arrive
\emph{during the period} and are discovered at time $T_i$ is

\[ c^0_i := \int_{T\imo}^{T_i} m(t) \bF(T_i-t) S(T_i-t)\,dt. \]
Some existing mortality estimators (see \Secs{ss:spec}{ss:comp}) embody the
assumption that all carcasses that arrived prior to the previous search at
time $T\imo$ will have been removed by scavengers or discovered and removed
by an FT in that earlier search, leaving none to ``bleed through'' from
earlier periods to be removed or discovered in the current search at time
$T_i$.  Under that assumption, $c^0_i$ would be the expected count
$\E[C_i]$.  Other mortality estimators are based on a different
assumption--- that undiscovered and unremoved carcasses from earlier
periods remain discoverable, so that $C_i$ may include both ``new''
carcasses from the current period and ``old'' ones that arrived during
earlier periods.  For $k\ge 1$ the expected number discovered at time $T_i$
that arrived during the $k$th previous period but were undiscovered in $k$
previous searches would be

\[ c^k_i := \int_{\Tim{k{-}1}}^{\Tim k} m(t) \bF(T_i-t) S(T_i-t)
          \prod_{0< n\le k}\big[1-S(\Tim n-t)\big]\,dt \]
and the total expected carcass count for the $i$th search would be
$\E[C_i]=c_i:=\sum_{k\ge0} c^k_i$.

Evidence (see \Sec{s:data}) suggests that \emph{both} the assumption that
all carcasses bleed through for later discovery, and the assumption that
none do, are wrong.  We here introduce an intermediate possibility: that
some fraction $0\le\bleed\le1$ do bleed through at each search, leading to
expected carcass count

\begin{align}
\E[C_i] = c_i :=
       \sum_{k=0}^\infty \bleed^k \int_{\Tim{k{-}1}}^{\Tim k}
       m(t) \bF(T_i-t) S(T_i-t)
       \prod_{0< n\le k} \big[1-S(\Tim n-t)\big]\,dt.\label{e:ci}
\end{align}
For slowly-varying $m(t)\approx m$, this leads to a maximum likelihood
estimate for the mean total mortality $m_i=\int_{T\imo}^{T_i}m(t)dt\approx
mI_i$ in period $(T\imo,T_i]$ of

\begin{subequations}\label{e:MRst}
\begin{align}
            \Mst_i &:= C_i/\Rsti,\label{e:Mst}
\end{align}
the carcass count $C_i$ inflated by a factor $1/\Rsti$ given by the inverse
of the ``reduction factor''

\begin{align}
    \Rsti &:= \frac1{I_i}
       \sum_{k=0}^\infty \bleed^k \int_{\Tim{k{-}1}}^{\Tim k}
       \bF(T_i-t) S(T_i-t) \prod_{0< n\le k}
       \big[1-S(\Tim n-t)\big]\,dt\label{e:Rst}
\end{align}
\end{subequations}
(so-called because on average the count $C_i\approx M_i \Rsti$ will be the
mortality $M_i$ reduced by the factor $\Rsti$).  For similar search
intervals $I_i\approx I$, the $k$th term in this sum for $k\ge1$ represents
carcasses that arrived between $kI$ days and $(k+1)I$ days before the end
of this search period, were unremoved by scavengers over that entire
period, were undiscovered and yet remained discoverable in $k$ consecutive
searches, and were finally discovered at time $T_i$.  This will be a rare
event unless $kI$ is quite small, so only a few terms of this sum are
typically sufficient to achieve accuracy within a few percent.  Simple
approximations and truncation error bounds for them are given in
\Sec{ss:dimin}. 

\citet{Shoe:2004} describes as \emph{periodic} those estimators (including
his own) based on the premise that all the undiscovered and unremoved
carcasses remain discoverable, and the assumption that consecutive periods
are similar.  Our proposed estimator, intermediate between the periodic
ones that assume 100\% bleed-through and the aperiodic ones that assume
0\%, might be described as \emph{partially-periodic}.

\subsection{Special Cases \& Previous Estimators}\label{ss:spec}
Before turning to the general case, consider first the simple situation
with constant removal rate (or \emph{hazard}) $[-\bF'/\bF](\tau)\equiv r$
and constant search proficiency $S(\tau)\equiv s$.  Under this assumption
that the scavenger removal rate and FT discover probabilities do not depend
on carcass age $\tau$, the removal times must follow the exponential
distribution $\tau\sim \Ex(r)$ with survival function
$\bF(t):=\P[\tau>t] =\exp(-rt)$ for $t>0$ and mean removal time
$\hat t:=\E[\tau] = 1/r$.  In that case, for constant inter-search
intervals $I_i\equiv I$, the reduction factor \eqref{e:Rst} simplifies to a
geometric series,

\begin{align}
    \Rsti &:= \frac1{I}
       \sum_{k=0}^\infty \bleed^k \int_{(i-k-1)I}^{(i-k)I}
       \exp\big(-r(i I-t)\big) s (1-s)^k\,dt\notag\\
         &= \frac{s~[e^{rI}-1]}
                 {rI[e^{rI}-\bleed(1-s)]}
          = \frac{s~\hat t~[e^{I/\hat t}-1]}
                 {I[e^{I/\hat t}-\bleed(1-s)]}.\label{e:Rst-sc}
\end{align}
In the case of zero bleed-through, $\bleed=0$ and \eqref{e:Rst-sc} leads to
the estimator

\begin{subequations}
\begin{align}
\hat M_i^P&= \frac{I\,C_i}{s~\hat t~[1-e^{-I/\hat t}]},\label{e:Poll}
\end{align}
that introduced by \citet{Poll:2007} (under exponentially-distributed
persistence).

\citet{Huso:2011} introduced a similar estimator $\hat M^H$ that differs in
replacing the term $[1-e^{-I/\hat t}]$ by
$\min\big(0.99, [1-e^{-I/\hat t}]\big)$.  The two are identical whenever
(as usual) search intervals $I$ are shorter than the mean removal times
$\hat t$ times a factor of $\log100\approx4.6$, for then
$[1-e^{-I/\hat t}]<0.99$ (otherwise Huso's estimator $M_i^H$ is up to 1\%
higher than Pollock's $M_i^P$).

In the case of full bleed-through, $\bleed=1$ and \eqref{e:Rst-sc} gives
the ``periodic'' estimator introduced by \citet{Shoe:2004},

\begin{align}
\hat M_i^S &= \frac{I\,C_i}{s~\hat t}
            \bet{\frac{e^{I/\hat t}-1+s}{e^{I/\hat t}-1}}.\label{e:Shoe}
\end{align}

Finally, setting $\bleed=1/(1-s)$ gives

\begin{align}
\hat M_i^E &= \frac{I\,C_i}{s~\hat t},\label{e:EJ}
\end{align}
\end{subequations}
the steady-state estimator introduced by \citet{Eric:Stric:etal:1998}.

\subsection{Comparing Current Estimators}\label{ss:comp}
All four of the estimators $\hat M_i^E$, $\hat M_i^S$, $\hat M_i^P$, and
$\hat M_i^H$ are special cases\footnote{For unusually long search intervals
  $I_i>4.6\hat t$ then $\hat M^H$ is up to 1\% higher than special case
  $\hat M^P$ of $\Mst$.  Also \citeauthor{Poll:2007}'s estimator $\hat M^P$
  is not limited to exponentially-distributed removal times $\tau$ with
  constant removal rate $r=1/\hat t$, although the method commonly used to
  estimate $\hat t$ \citep [\S2.6 \& \S3.2] {PSE:2008} is the MLE for that
  case and is badly biased for heavier-tailed distributions.}
%
%
%
%
of \eqref{e:Rst-sc}, for specific values of $\bleed$.  Always

\begin{align}
\big[1-e^{-(4.6\wedge I/\hat t)}\big] \hat M_i^H =
\hat M_i^E < \hat M_i^S < \hat M_i^P \le \hat M_i^H, \label{e:comp}
\end{align}
so all four estimators are within 5\% if $I > 3\hat t$ and within $58\%$
for $I > \hat t$.  Under the assumptions of constant removal rate
$-\bF'/\bF\equiv r$ and constant searcher proficiency $S\equiv s$, the
proposed new estimator $\Mst$ of \eqref{e:MRst} also lies in the interval
$[\hat M^E_i, \hat M^H_i]$ for any $0\le\bleed\le1$.

Differences among the estimators will be substantial for shorter search
intervals, however.  For example, for search intervals substantially
shorter than the mean scavenger removal time, $I\ll\hat t$ and so

\[ \hat M^H_i \ge \hat M^P_i > (\hat t/I) \hat M^E_i \gg \hat M^E_i ,\]
and it will be important to assess bleed-through rate $\bleed$ accurately.
And, if the assumptions of constant removal rates and search proficiencies
are incorrect, then the estimators may agree with each other but all be
badly biased.

\section{Variable Search Proficiency and Removal Rates}\label{s:vary}
Both the assumptions of constant removal rate and of constant search
proficiency, irrespective of carcass age, appear inconsistent with the
observations presented in \Sec{s:data}.  In this section we show how to go
beyond those assumptions.
\subsection{Diminishing Proficiency}\label{ss:dimin}
For many data sets the search proficiency $S(t)$ appears to diminish with
increasing carcass age $t$.  In \Sec{s:data} it is shown that the data are
fit well by an exponentially decreasing success rate

\begin{align}
 S(t)=\exp\big(-a-bt\big) \label{e:disc}
\end{align}
for parameters $a,b\ge0$ (logistic models gave very similar results).  With
this modeling choice, and for equal search intervals $I_i=I$ (say, with
searches at times $T_i=iI$), the ACME estimator $\Mst_i$ and reduction
factor $\Rsti$ of \eqref{e:MRst} take the form

\begin{subequations}\label{e:MRst2}
\begin{align}
            \Mst_i &:= C_i/\Rsti,\label{e:Mst2}
\end{align}
with

\begin{align}
    \Rsti
    &:= \sum_{k=0}^\infty \bleed^k \int_{k}^{k+1}
        \bF\big(xI\big)~  e^{-a-xbI}
        \prod_{0< n\le k} \Big[1-e^{-a-(x-n)bI} \Big]\,dx
        =\sum_{k=0}^\infty\Tst_k,\label{e:Rst2}
\end{align}
whose $k$th term $\Tst_k$ represents the fraction of carcasses that arrived
in the search period ending at $\Tim k$ that are discovered at time $T_i$.
Of particular importance (see \Sec{s:interval}) is the first of these

\begin{align}
\Tst_0 &= \int_0^1 \bF\big(xI\big)~  e^{-a-xbI}\,dx,\label{e:Tst0}
\end{align}\end{subequations}
the fraction of carcasses discovered at the search ending the interval in
which they arrived.  Each $\Tst_k$ is expressible as the sum of $2^k$ terms
of the form

\begin{align}
 \Qst_{kmn}
      &:= \bleed^k (-1)^{m+1}
        \int_0^1 \bF\big((k+x)I\big)
                 e^{-m\,(a+bI\,x)-n\,bI}\,dx\label{e:Qst}
\end{align}
for suitable nonnegative integers $m,n$ that can be enumerated recursively:
beginning with $(k,m,n)=(0,1,0)$, each entry $(k,m,n)$ generates at the
next level $(k+1,m,n+1)$ and $(k+1,m+1,n+k+1)$.  The first few terms are

\vspace*{-1mm}\begin{align}
  \Tst_0 &= \Qst_{010}\label{e:T0}\\
  \Tst_1 &= \Qst_{111} + \Qst_{121}\notag\\
  \Tst_2 &= \Qst_{212} + \Qst_{223} + \Qst_{222} + \Qst_{233}\notag\\
  \Tst_3 &= \Qst_{313}+\Qst_{325}+\Qst_{324}+\Qst_{336}
  +\Qst_{323}+\Qst_{335}+\Qst_{334}+\Qst_{346}\notag
\end{align}
The truncation error from using only the first $N$ terms $0\le k<N$ of
the infinite sum in \eqref{e:Rst2} is bounded by

\begin{align}
0\le \text{Truncation Error}
   \le \bleed^N\bF(N\,I)\bet{\frac{e^{-a-N\,bI}}
    {bI \vee (1-\bleed e^{-bI})} \wedge (1-e^{-a})^N}.
\label{e:trunc}\end{align}
For the examples presented in \Sec{s:data}, the truncation error
bound is about 1\% of $\Rsti$ with $N=3$ terms, and about $0.1\%$
with $N=5$ terms.

\subsection{Persistence Distributions}\label{ss:persist}
\citet {Bisp:Bern:etal:2013a,Bisp:Bern:etal:2013b} found (and we verify in
\Sec{ss:rem} below) that log normal, log logistic, and Weibull
distributions with decreasing hazard functions all fit empirical
persistence data quite well, and that exponential distributions did not.
Here we take the Weibull distribution, parametrized in the form

\begin{align}\label{e:rem}
 \bF(t) := \P[\tau>t] = \exp\big(-(\rho t)^\alpha\big),\qquad t>0
\end{align}
for rate $\rho>0$ (in units of $\dy^{-1}$) and unitless shape parameter
$\alpha>0$.  For this distribution the key quantities $\Qst_{kmn}$ from
\eqref{e:Qst} needed to compute $\Rsti$ are

\begin{align*}
 \Qst_{kmn}
      &:= \bleed^k (-1)^{m+1}
        \int_0^1 \exp\big(-(\rho(k+x)I)^\alpha
                 -m\,(a+bI\,x)-n\,bI\big)\,dx,
\end{align*}
easily evaluated numerically using Simpson's quadrature rule or, for the
particular values of $\alpha=\half$ and $\alpha=1$, available explicitly in
closed form:

\begin{align}
 \Qst_{kmn} &= \frac{2\bleed^k(-1)^{m+1}}{\rho I}
        \exp\big(-ma+(mk-n)bI+\rho/4mb\big)\tag{$\alpha=\half$}\\
    &\times \Bigg\{\frac{\rho}{2mb}\Big[e^{-mb(\sqrt{kI\rho}+\rho/2mb)^2}
           -e^{-mb (\sqrt{\kpo I\rho}+\rho/2mb)^2}\Big]\notag\\
    &\quad +2\sqrt{\frac{\pi\rho}{mb}}\Big[
      \Phi\Big(\sqrt{{2mb}/{\rho}}
        \big(\sqrt{\rho k    I}+\frac{\rho}{2mb}\big)\Big)-
      \Phi\Big(\sqrt{{2mb}/{\rho}}
        \big(\sqrt{\rho \kpo I}+\frac{\rho}{2mb}\big)\Big)
        \Big] \Bigg\}\notag\\
 \Qst_{kmn} &=
  \frac{(-1)^{m+1}e^{-m(a+bI)}} {(\rho+mb)I} \big[ 1- e^{-(\rho+mb)I} \big]
        \big(\bleed e^{-rI}\big)^k
 \tag{$\alpha=1$}
\end{align}
where $\Phi(z)$ denotes the CDF for the standard $\No(0,1)$ normal distribution.

\section{Mortality Estimates}\label{s:interval}
Point estimates like $\Mst$ of \eqref{e:Mst} and \eqref{e:Mst2} are more
informative when accompanied by some measure of their uncertainty.  For
example, \citet {Eric:Stric:etal:1998} recommend reporting 50\% and 90\%
interval estimates for mortality.


\subsection{Interval Estimates for Mean Mortality $m_i$}\label{ss:mm}

In this section we will find interval estimates for the \emph{mean} daily
mortality rate $m_i$ based on observed carcass counts $C_i$.  Such an
estimate is given by a pair of functions $\mathtt{lo(c)}$ and
$\mathtt{hi(c)}$ with the property that

\[ \P\big[m_i \in \mathtt{[lo(C_i),~hi(C_i)]}~\big] \ge \gamma \]
for specified $\gamma$ (such as $0.5$ or $0.9$, per \citet
{Eric:Stric:etal:1998}).  The common symmetric choice is to arrange that
$\P\big[m_i < \mathtt{lo(C_i)}\big]$ and
$\P\big[m_i > \mathtt{hi(C_i)}\big]$ are each below $(1{-}\gamma)/2$.
Frequently in practice however mortality is low enough (or removal is rapid
enough) that observed counts as low as zero or one are common
\citep{Huso:Dalt:etal:2014}, motivating interest in \emph{one-sided}
interval estimates with $\mathtt{lo(c)}\equiv 0$ and
$\P\big[0\le m_i \le \mathtt{hi(C_i)}\big]\ge\gamma$.  A third option is to
find the \emph{shortest} interval that captures $m$ with probability at
least $\gamma$.

Under the model introduced in \Secs {s:model} {s:vary} the mortality $M_i$
in the $i$th search period $(\Tim1,T_i]$ has a Poisson distribution whose
mean is the product $m_iI_i$ of the average daily mortality in that
period $m_i$ and the search period length $I_i=(T_i-\Tim1)$.  If these
  rates and lengths are nearly constant (say, $m_i\approx m$ and
  $I_i\approx I$) over the period during which all the carcasses found at
  time $T_i$ arrived, and if the model parameters determining the reduction
  factor $\Rsti$ of \Eqn{e:Rst2} are nearly constant, then the conditional
  (given $m$) distribution of $C_i$ is

\[ C_i\mid m \sim \Po\big(\Rsti mI). \]
  With conjugate Gamma prior distribution $m\sim\Ga(\malp,\mbet)$ (more on
  this below), the marginal distribution of carcass counts is negative binomial

\begin{align}
  C_i\sim\NB(\malp,\mbet/(\mbet+\Rsti I)) \label{e:Ci-marg}
\end{align}
and the posterior distribution for $m$ given $C_i$ is again Gamma but with
new parameters:

\begin{align}
        m\mid C_i &\sim \Ga(\malp+C_i, \mbet+\Rsti I).\label{m|C}
\end{align}
The Objective Bayes reference prior distribution \citep{Berg:Bern:Sun:2009}
for $m$, expressing no available prior or extrinsic information about it,
is the improper $m\sim m^{-\half}$, the limiting case of the Gamma
distribution with $\malp=\half$ and $\mbet=0$.  An alternative to Objective
Bayes is to follow an Empirical Bayes approach \citep{Robb:1955, Case:1985}
using the evidence about $m$ reflected by previous observations of
$\{C_i\}\iid\NB(\malp,\mbet/(\mbet+\Rsti I))$ (typically this leads to
shorter intervals, since they reflect more evidence about the average
mortality rate $m$).  It proceeds by making (often Maximum Likelihood)
estimates $\hat\malp$ and $\hat\mbet$ of the parameters, and basing
interval estimates for $m$ on these.

The resulting posterior $\gamma=50\%$ or $\gamma=90\%$ Credible Interval
estimates for $m$ are of the form $\big[\mathtt{lo(C_i),~~hi(C_i)}\big]$
with the functions $\mathtt{lo(c)}$ and $\mathtt{hi(c)}$ given by one
of:\par
\begin{tabular}{lLL}
Objective Bayes, One-Sided:&
  \mathtt{lo(c) = 0}\\
& \mathtt{hi(c) = qgamma(\gamma,~c+\half,~\Rsti I)}\\
Objective Bayes, Symmetric:&
  \mathtt{lo(c) = qgamma((1-\gamma)/2,~c+\half,~\Rsti I)}\\
& \mathtt{hi(c) = qgamma((1+\gamma)/2,~c+\half,~\Rsti I)}\\
Empirical Bayes, One-Sided:&
  \mathtt{lo(c) = 0}\\
& \mathtt{hi(c) = qgamma(\gamma, \hat\malp+c, \hat\mbet+\Rsti I)}\\
Empirical Bayes, Symmetric:&
  \mathtt{lo(c) = qgamma((1-\gamma)/2,~\hat\malp+c,~\hat\mbet+\Rsti I)}\\
& \mathtt{hi(c) = qgamma((1+\gamma)/2,~\hat\malp+c,~\hat\mbet+\Rsti I)}\\
\end{tabular}
\par\smallskip\noindent where $\mathtt{qgamma(x,a,b)}$ \citep{R:2014}
denotes the quantile function (inverse CDF) for the Gamma distribution.  If
the mortality rate $m(t)$ varies slowly enough that it may be considered
constant over a longer period of time including some $n\ge2$ search
intervals of total length $I_+:=(T_i-\Tim n)$, then the total number of
carcasses $C_+:=\sum C_i$ found in the $n$ searches will again have a
Poisson conditional distribution $C_+\mid m\sim \Po( \Rsti I_+ m)$ and a
Negative Binomial marginal distribution
$C_+\sim \NB(\malp, \mbet/ (\mbet+\Rsti I_+))$, and the posterior for $m$
will again be Gamma, $m\mid C_+ \sim \Ga(\malp+C_+, \mbet+\Rsti I_+)$.
Quantiles of this Gamma distribution will determine Credible Intervals for
$m$ that will be narrower by approximately a factor of $\sqrt n$ than those
of \eqref{m|C}, and so will specify $m$ to higher precision.  The
assumption of near-constancy of $m$ and the model parameters determining
$\Rsti$ would be violated for periods long enough to include changes in
season, vegetation, or migratory patterns.

\subsection{Interval Estimates for Mortality $M_i$}\label{ss:M-int}
In this section we find interval estimates for the number $M_i$ of
carcasses that arrived in the interval $(T\imo,T_i]$ based on the observed
carcass count $C_i$.  These will be wider than the intervals for $m_i$ of
\Sec{ss:mm} because the aleatoric uncertainty and variability of mortality
events typically exceeds the epistemic uncertainty about parameter values.

In general the $C_i$ carcasses discovered in the search at time $T_i$ may
include both some of the $M_i$ carcasses that arrived during the period as
well as some of those that arrived in earlier periods.
Thus there is no way of making meaningful interval estimates about $M_i$
from $C_i$ alone, without making some assumptions about either the
$\{M_j\}$ for $j<i$, \ie, about mortality in the recent past, or about the
absence of bleed-through.

\subsubsection{Classical Confidence Intervals ($\bleed=0$
  only)}\label{sss:class}
If, despite the evidence in \Sec{s:data}, one assumes that no carcasses
from earlier periods are ever discovered, \ie, if $\bleed=0$, then
$C_i \sim \Bi(M_i, \Rsti)$ and classical Confidence Interval estimates are
available for this binomial model without concern for mortality in earlier
periods.  For example, a 90\% one-sided classical confidence interval for
$M_i$ would be $[\mathtt{C_i, hi(C_i)}]$, where

\[ \mathtt{hi(c)=}\inf\{M\ge c:~\pbinom{c,~M,~\Rsti}\le 0.10\}
\]
where $\pbinom{x,n,p}$ \citep{R:2014} denotes the CDF for the Binomial
distribution. 

\subsubsection{Objective Bayes Credible Intervals (any $\bleed$)}\label{sss:OB}
No simple classical confidence intervals for $M_i$ are available for the
more realistic situation of $\bleed>0$.  Again, however, Objective Bayes
and Empirical Bayes credible intervals may be constructed for $M_i$ based
on the model of \Secs {s:model} {s:vary}.  Both Objective and Empirical
Bayes posterior distribution for $M_i$, given $C_i$, are derived in
Appendix \ref{ss:post-mort} and presented as

\begin{align}\label{e:M|C}
  \P[M_i = M\mid C_i=C] = c \times {}_2F_1(-C,-M;\malp-C-M;-z)
\end{align}
with $\malp=\half$ for Objective Bayes or $\malp=\hat\malp$ for Empirical
Bayes, for specified quantities $c$ and $z$ given in \Eqns{e:M|C-obj}
{e:M|C-emp}, respectively, as explicit functions of $\Rsti$ and $\Tst_0$
from \Eqns{e:Rst2}{e:Tst0} (here ${}_2F_1(a,b;c;z)$ denotes Gauss'
hypergeometric function \citep [\S15]{DLMF:2014}).  Setting
$\mathtt{p(m|c)}:=\P[M=m\mid C=c]$ from \eqref{e:M|C}, credible intervals
for $M$ are

\begin{align}[\lo{C_i},\hi{C_i}]\label{e:M-int}\end{align}
with
\par\centerline{\begin{tabular}{lLLL}
One-sided:&\mathtt{lo(c)=}0\\
          &\mathtt{hi(c)=}\min\{M:~\sum_{m\le M}
           \mathtt{p(m|c)} \ge \gamma\}\\
Symmetric:&\mathtt{lo(c)=}\max\{M:~\sum_{m\le M}
           \mathtt{p(m|c)} \le (1-\gamma)/2\}\\
          &\mathtt{hi(c)=}\min\{M:~\sum_{m\le M}
           \mathtt{p(m|c)} \ge (1+\gamma)/2\}
\end{tabular}}
\noindent
while Highest Posterior Density or HPD intervals \citep [the shortest
possible intervals with coverage probability $\gamma$, see] [\S2.3]
{Gelm:Carl:Ster:Rubi:2009} for $M$ upon observing $C_i=c$ are available by
sorting the values $\{\mathtt{p(m|c):}\,m\ge0\}$ in decreasing order and
identifying the smallest collection whose sum exceeds $\gamma$.  Some of
these distributions and intervals are shown in \Fig{f:post}.  Similar
Empirical Bayes results are available from \Eqns {e:M|C} {e:M|C-emp} with
estimated hyperparameters $\hat\malp, \hat\mbet$.

In the absence of bleed-through (\ie, $\bleed=0$) all found carcasses are
``new'' so necessarily $M_i\ge C_i$.  It is shown in \Sec{sss:B=0} that the
number $(M_i-C_i)$ of undiscovered carcasses then has the Negative Binomial
distribution
$(M_i-C_i)\mid C_i\sim\NB\big(\malp+C, (\mbet+\Rsti I)/(\mbet +I)\big)$, so

\begin{align}\label{e:M|C-B=0}
 \P[M_i=M\mid C_i=C] = \frac{\Gamma(\malp+M)}{\Gamma(\malp+C)~(M-C)!}
(\Rsti+\mbet/I)^{\malp+C}(1-\Rsti)^{M-C}(1+\mbet/I)^{-\malp-M}
\end{align}
from which credible intervals for $M_i$ are available.  For example, the
one-sided Objective Bayes interval is $[\lo{C_i},\hi{C_i}]$ with

\begin{subequations}\label{e:B0}\begin{align}
\lo c &= c \qquad\qquad \hi c = c+ \qnbinom{\gamma, c+ 1/2, \Rsti}
\label{e:B0half}\end{align}
\noindent
where $\mathtt{qninom(p, alpha, prob)}$ \citep{R:2014} denotes the quantile
function for the negative binomial distribution.  HPD regions are available
with a search.

A more direct and less model-dependent Bayesian approach to finding the
conditional distribution of $M$ given $C$ would be to begin with an
improper uniform prior distribution for $M$ on the nonnegative integers
$\{0,1,\dots\}$.  The posterior distribution of the unobserved carcass
count $(M-C)$, after observing $C\sim\Bi(M,\Rst)$, then has the negative
binomial distribution $(M_i-C_i)\mid C_i\sim \NB(C_i+1,\Rsti)$, leading to
very similar one-sided intervals with 

\begin{align}
\lo c &= c \qquad\qquad \hi c = c+ \qnbinom{\gamma, c+ 1, \Rsti}.
\label{e:B01}\end{align}\end{subequations}

\section{Results from Altamont}\label{s:data}
\citet{WaHi:Newm:Etal:2012} report on data taken from January 7 to April 30
of 2011 in the Altamont Pass Wind Resource Area in a study of the removal
and discovery rates of aging bird and bat carcasses.  One hundred and ten
bird carcasses (predominantly brown-headed cowbirds, \emph{Molothrus Ater},
with AOU code BHCO \citep{Pyle:DeSa:2014}) and 78 bat carcasses of
disparate species were placed by Project Field Managers (PFMs), who then
checked every few days to confirm whether or not each carcass remained in
place.  Field Technicians (FTs) would search for carcasses at approximately
one week intervals, noting the species and location of those they
discovered but not disturbing or removing them.  Successive searches were
conducted by different FTs who were unaware of any earlier carcass
discoveries.  This ``integrated detection trial'' or IDT design
\citep[Chap.\,2] {WaHi:Newm:Etal:2012} afforded the possibility of
exploring how removal rates and discovery probabilities may change over
time.

\subsection{Removal by scavengers}\label{ss:rem}
\begin{figure}[!htp]
\begin{center}
\includegraphics[width=0.95\textwidth,trim=0 30 0 10]{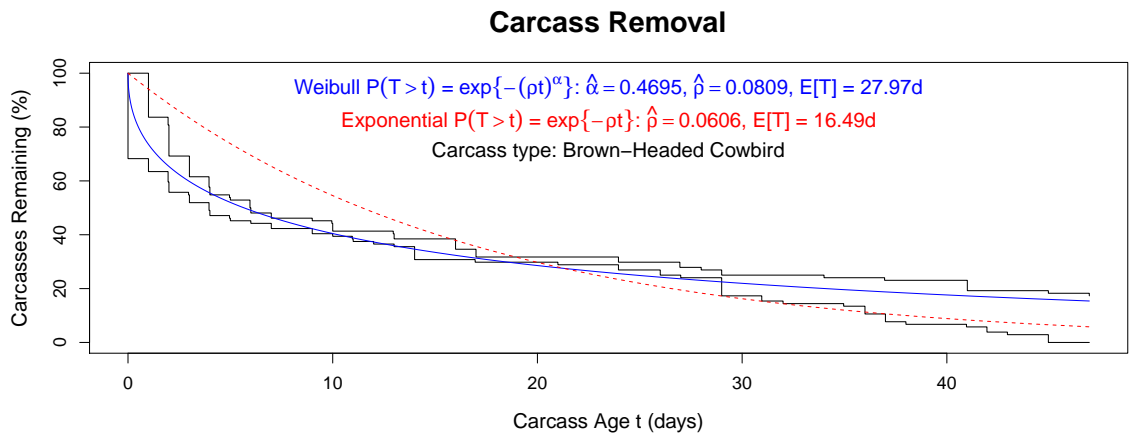}
\caption{\label{f:bhco-scav}Empirical survival function showing removal
  of brown-headed cowbirds by scavengers (solid black stair-step lines),
  along with best Weibull distribution fit (solid blue line) and best
  Exponential distribution fit (dashed red line).  Note Weibull fits
  well while Exponential does not.}
\end{center}
\end{figure}

\Fig{f:bhco-scav} illustrates the removal of brown-headed cowbird
carcasses by scavengers.
Removals are interval censored: we only observe the times of the last
recorded discovery of a carcass's presence and the first of its absence.
Thus the empirical survival function in \Fig{f:bhco-scav} consists of two
black stair-step curves based on the earliest and latest possible times of
removal consistent with the observations.  The best Weibull distribution
fit (see \Sec{sss:param-rem} for derivation of likelihood function
\eqref{e:nllh-rem} and MLEs),

\[ \P[\tau > t] = \exp\big(-(\rho t)^\alpha\big),\qquad
   \hat\alpha=0.4695,\quad \hat\rho = 0.0809\dy^{-1}
\]
is illustrated with the solid blue curve.  Its mean of $\E[\tau] =27.97\dy$
is nearly twice that ($16.49\dy$) of the best exponential distribution fit,
shown as a dashed red line.  The exponential distribution model
underestimates early removal rates and overestimates later ones.  The
estimated shape parameter $\hat\alpha=0.4695$ is $9.8$ standard errors away
from the value $\alpha=1$ for the exponential distribution, making the
exponential distribution and its assumption of constant removal rates
entirely untenable.  Best fits with log normal and log logistic were nearly
indistinguishable from Weibull, so we present only Weibull results here.

\subsection{Search Proficiency and Bleed-through}\label{ss:srch}
\begin{figure}[!htp]
\begin{center}
\includegraphics[width=0.95\textwidth,trim=0 30 0 10]{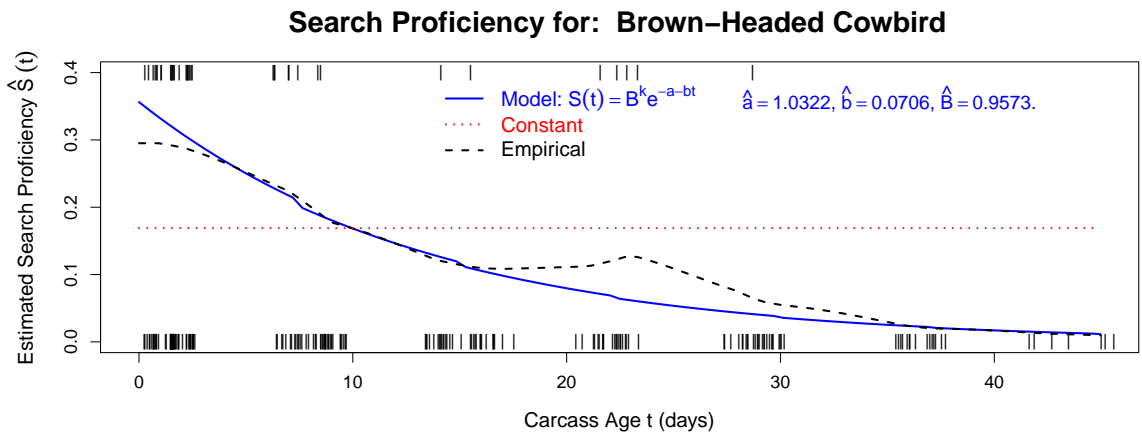}
\caption{\label{f:bhco-srch}Empirical plot of search proficiency (dashed
  black line) for brown-headed cowbirds, along with best fit
  exponentially-decreasing curve (solid blue line) and best fit of a
  constant proficiency (dashed red line).  Search successes (and failures)
  are shown as whiskers at the top (and bottom) of the plot, respectively.
  Note diminishing proficiency model fits data well while constant
  proficiency does not.}
\end{center}
\end{figure}

\Fig{f:bhco-srch} illustrates the search process by FTs.
Short vertical dashes at the top and bottom of the plot indicate the times
of successful and unsuccessful searches, respectively.  Dashed black curve
indicates a nonparametric estimator of time-dependent search proficiency, a
moving-average double-exponential window estimator with width of 5\dy.
Proficiency exceeds $30\%$ initially, but falls off at about $7\%\dy^{-1}$.

Solid blue line shows best exponentially-decreasing fit, based on MLEs
$\hat\alpha=0.4695$, $\hat\rho=0.0808$, and $\hat\bleed=0.9573$ found by
minimizing the negative log likelihood of \Eqn{e:disc} (see
\Sec{sss:param-disc}).  Dotted red line shows best constant-proficiency
fit.

The deviance between the proposed model and the constant-proficiency model,
a sub-model with $b=0$ and $\bleed=1$, is $D=22.63$.  By Wilks' theorem
\citep{Wilk:1938} this would have approximately a $\chi^2_2$ distribution
with two degrees of freedom if the constant-rate model were correct,
evidently an entirely untenable supposition with $P$-value about $10^{-5}$.

Carcasses were later discovered after an initial miss 9 times in this
study, and after some earlier miss 12 times, confirming that some
bleed-through occurred.  Estimated bleed-through rate is
$\hat\bleed=95.73\%$.  Evidence against full bleed-through $\bleed=1$ is
not strong enough to reject that possibility.

\subsection{Mortality Estimation at Altamont}\label{ss:mort-alt}
With the parameter estimates

\[ \hat\alpha=0.4695\qquad
   \hat\rho=0.0809\dy^{-1}\qquad
   \hat a=1.0322\qquad
   \hat b=0.0706\dy^{-1}\qquad
   \hat\bleed=0.9573
\]
for the Weibull removal distribution ($\alpha,\rho$), exponentially falling
search proficiency ($a,b$), and bleed-through rate ($\bleed$) (see \Sec
{ss:params}), we can use \eqref{e:Tst0} and a five-term approximation to
\eqref{e:Rst2} to evaluate the Reduction Factor $\Rsti$ for future searches
at $7\dy$ intervals and the fraction of ``new'' carcasses $\Tst_0$ found in
each search:

\begin{align*}
    \Rsti &= 0.2496\qquad\qquad \Tst_0 = 0.1740.
\end{align*}
This suggests that about a quarter of the carcasses are discovered
eventually, $17\%$ in the first search after arrival and the rest following
bleed-through.  This leads to the ACME adjusted mortality estimate

\begin{align*}
\Mst_i & = C_i/\Rsti = 4.01 \times C_i
\end{align*}
for a seven-day interval ending in a search at which $C_i$ brown-headed
cowbird carcasses are discovered.  From the same data and parameter
estimates we can find reduction factors for other possible search interval
lengths.  For example, $\Rsti=0.14$ and $\Mst_i=6.9\times C_i$ for
$I=14$-day searches, while $\Rsti=0.47$ and $\Mst_i=2.1\times C_i$ for
$I=2$-day searches and $\Rsti=0.57$, $\Mst_i=1.8\times C_i$ for daily
searches.

\begin{figure}\centering
\begin{tabular}{cc}
 \includegraphics[width=0.45\textwidth] {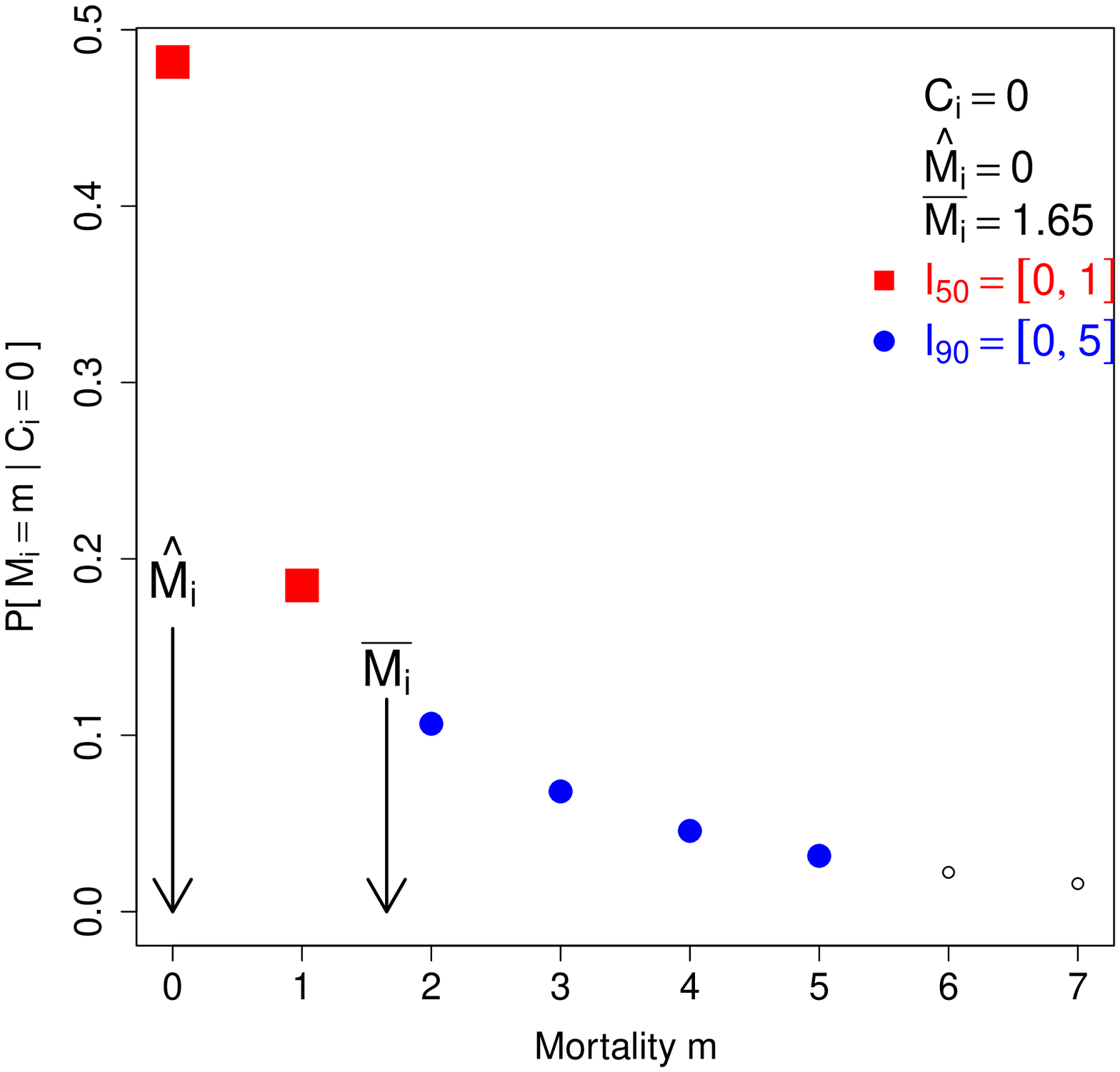} &
 \includegraphics[width=0.45\textwidth] {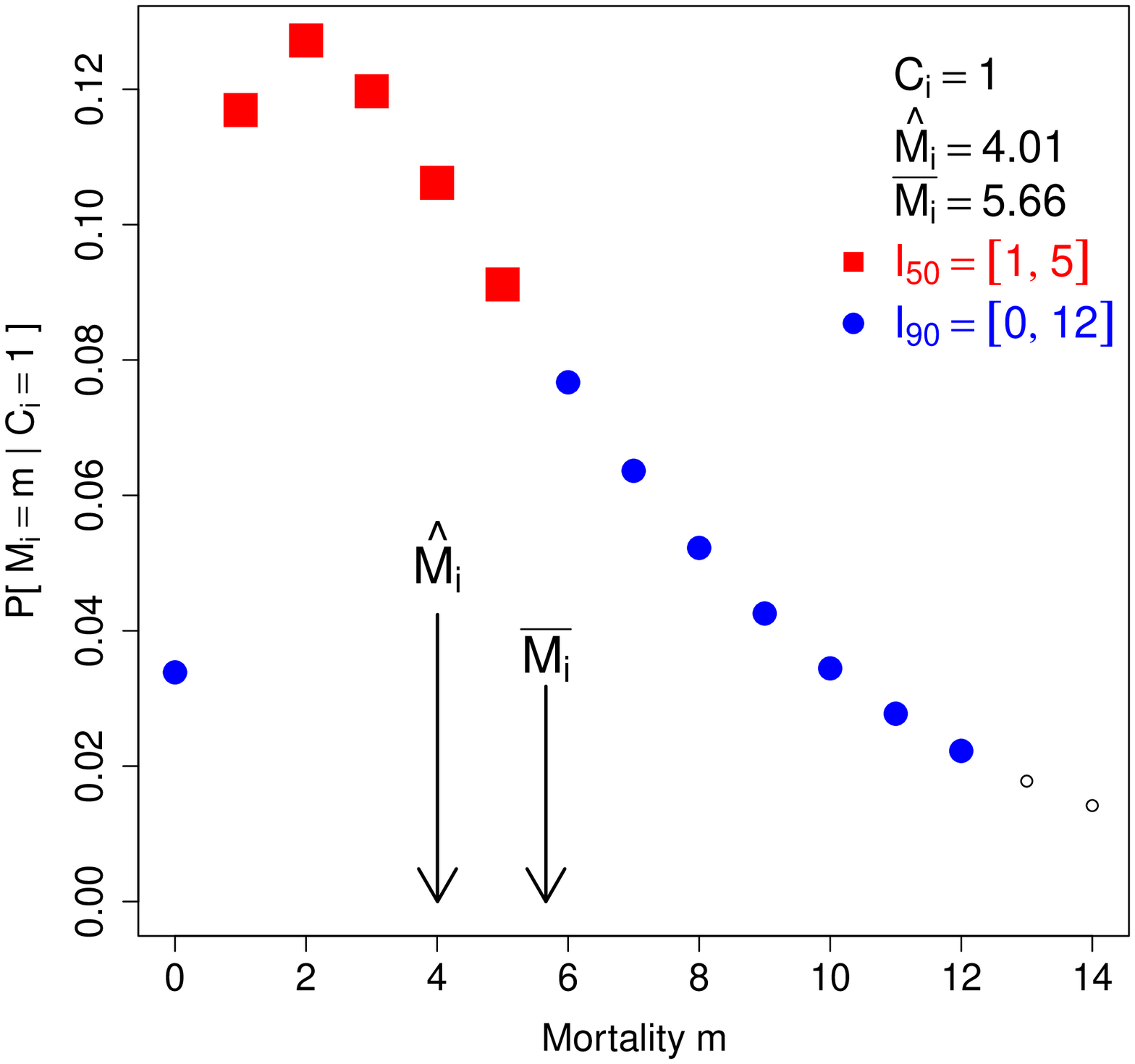}\\(a)&(b)\\
 \includegraphics[width=0.45\textwidth] {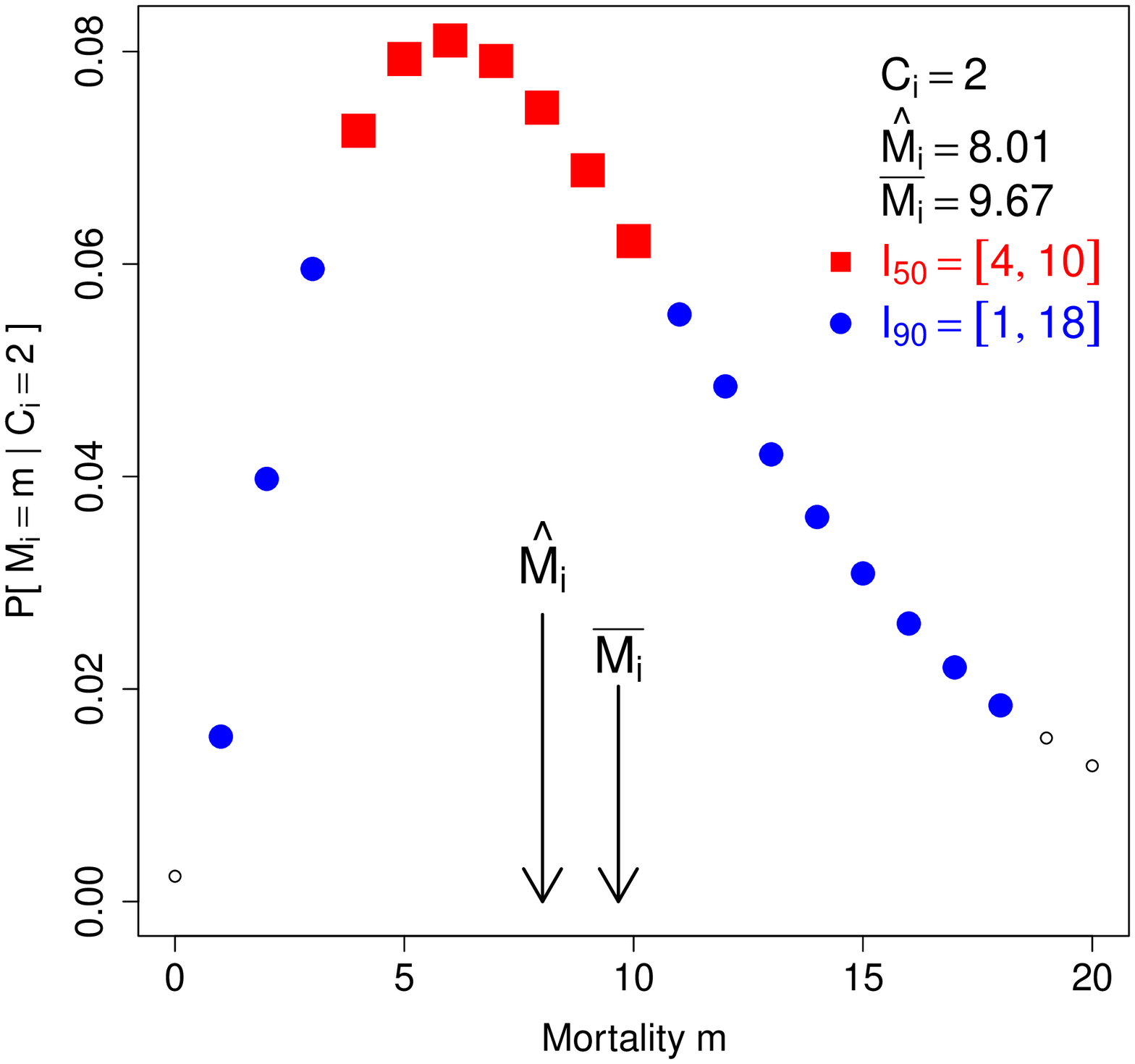} &
 \includegraphics[width=0.45\textwidth] {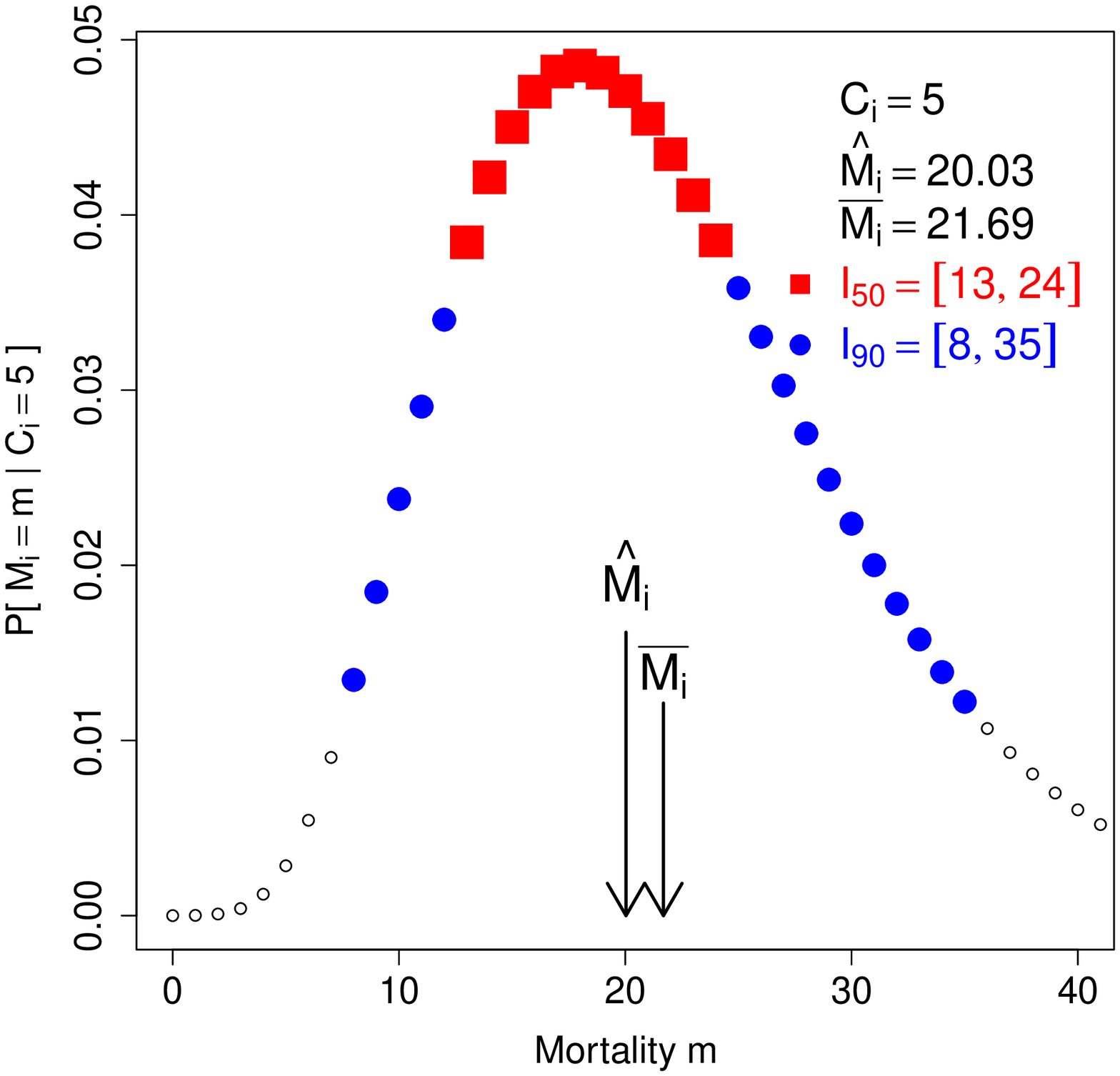}\\(c)&(d)
\end{tabular}
\caption{\label{f:post}Objective Bayes posterior distribution of mortality
  $M_i$ for brown-headed cowbirds using $7$-day search intervals for
  carcass counts $C_i=0,1,2,5$ in panels (a), (b), (c), (d), respectively,
  based on \Eqns{e:M|C} {e:M|C-obj}.  Large red squares show 50\% HPD
  credible intervals, filled blue disks show 90\% intervals.  Downward
  arrows indicate ACME estimates $\Mst_i = C_i/\Rsti$ and Objective Bayes
  posterior means $\overline M^\star_i$.}
\end{figure}

\Fig{f:post} shows Objective Bayes posterior distributions (see \Eqns
{e:M|C} {e:M|C-obj}) for the Brown Cowbird mortality $M_i$ at Altamont in a
7-day search period in which $C_i$ carcasses were discovered for a few
small values of $C_i$.  Also given in the figure legends are point
estimates $\Mst_i =C_i/\Rst$, posterior means $\overline M^\star_i$, and
50\% and 90\% Objective Bayes posterior HPD interval estimates $I_{50}$ and
$I_{90}$, derived in \Sec {ss:post-mort}.  These are also indicated in the
figure by vertical arrows at $\Mst_i$ and $\overline{M}^\star_i$ and by
large red squares and filled blue disks illustrating $I_{50}$ and $I_{90}$,
respectively.

\par

\section{Discussion}\label{s:disc}

Commonly-used existing estimators give similar results if search intervals
$I_i$ are much longer than the typical time $\hat t_i$ carcasses remain
unremoved by scavengers, but differ drastically for more frequent searches
because some of these estimators assume that undiscovered carcasses may
remain from one search period to the next and some do not.  Even when they
agree they may be biased by disregarding the diminishing removal rate (by
scavengers) and discovery proficiency (by Field Technicians) as carcasses
age.

This work presents a new estimator called ACME (an acronym for Avian and
Chiropteran Mortality Estimator) that includes many existing estimators as
special cases, but that extends them in three ways: it reflects diminishing
removal rates; it reflects decreasing discovery proficiency; and it allows
for an arbitrary rate of ``bleed-through'' of carcasses that arrived before
the current search period began.  It also includes interval (as well as
point) mortality estimates.

Mathematical formulas and computational methods are derived and presented
here for both the initial problem of estimating the model's five parameters
on the basis of field discovery trials, and the continuing problem of
constructing point and interval estimates for mortality on the basis of
these parameter estimates and subsequent observed carcass counts.

\subsection*{Data Accessibility}
A software package \texttt{acme} in the open-source \texttt{R} computer
environment \citep{R:2014} is available at \texttt{CRAN} for finding
maximum likelihood estimates of the model parameters and for evaluating the
ACME estimator $\Mst$, to make use of this estimator more accessible.  Data
used in preparation for this paper are included in that package.  A guide
to the design of integrated discovery trials suitable for supporting
inference about the diminishing rates of discovery and removal (often
unavailable from current discovery trial protocols) is also under
development.

\section*{Acknowledgments}\label{s:ack}
This work was primarily supported by the California Wind Energy
Association.  Additional support was provided by National Science
Foundation grants NSF DMS--1228317 and PHY--0941373 and by NASA AISR grant
NNX09AK60G.  Any opinions, findings, and conclusions or recommendations
expressed herein are those of the author and do not necessarily reflect the
views of CalWEA, the NSF, or NASA.  The author is grateful to William
Warren-Hicks and Brian Karas for access to data and to both them and to
Taber Allison, Regina Bispo, Jake Coleman, Daniel Dalthorp, Manuela Huso,
and James Newman for helpful conversations and insight.  After this work
was completed the author learned of independent related work by some others
\citep {Ette:2013, KoNi:Behr:etal:2015, Pero:Hine:Nich:etal:2013} with some
parallels to the current work.

\appendix
\section{Appendix: Computational Details}\label{s:app}

\subsection{Parameter Estimates}\label{ss:params}
In this section we construct maximum likelihood estimates from Integrated
Detection Trial (IDT) data for the five parameters
$(\alpha, \rho,~ a, b, ~\bleed)$ needed for the model of \Secs {s:model}
{s:vary} to support point estimates $\Mst_i := C_i/\Rsti$ of \Eqn{e:MRst2}
and interval estimates $[\lo{C_i},\hi{C_i}]$ of \Sec{ss:M-int} for
mortality $M_i$. 

\subsubsection{Removal}\label{sss:param-rem}
Persistence times in this model have the Weibull distribution \eqref{e:rem}
with $\P[\tau>t] = \exp\big(-(\rho t)^\alpha\big)$ for $t>0$, depending on
the two parameters $\alpha$ and $\rho$.  Carcass placement times $t_0$ are
known, but removal times $t_r$ (by scavengers) are generally not observed.
The data available from an IDT bearing on $(\alpha,\rho)$ from the $k$th
carcass consist of its placement time $t^k_0$, the last time
$t^k_p\ge t^k_0$ of its known presence from discovery by either a FT or
PFM, and the first time $t^k_a\ge t^k_p$ of its confirmed absence by a PFM
(or $t^k_a=\infty$ if it remains present throughout the trial).  The
negative log likelihood function on the basis of these interval-censored
data is

\begin{align}
\ell_\mathrm{rem}(\alpha,\rho)
   &= - \sum_{k}\log\set{e^{-[\rho(t^k_p -t^k_0)]^\alpha}
                       -e^{-[\rho(t^k_a -t^k_0)]^\alpha}}\notag\\
   &= \rho^\alpha\sum_k(t^k_p-t^k_0)^\alpha
      - \sum_{k}\log\set{1 -e^{\rho^\alpha
                 [(t^k_p-t^k_0)^\alpha-(t^k_a-t^k_0)^\alpha]}}.\label{e:nllh-rem}
\end{align}
The MLEs presented in \Sec{ss:rem} are the minimizing values
$(\hat\alpha,\hat\rho)$, easily found by a numeric search, along with
approximate standard errors from the inverse Hessian.

\subsubsection{Discovery}\label{sss:param-disc}
The probability of discovery of a $t$-day-old carcass present at an FT's
search is given in \eqref{e:disc} as $S(t)=\exp\big(-a-bt\big)$, depending
on the two parameters $(a,b)$.

Again denote by $t_0$ the placement time for a particular carcass (say, the
$k$th) and by $t_p$ the last time it is known to be present.  Let
$m_0:=\min\{n:T_n\ge t_0\}$ and $m^*:=\max\{n\ge m_0:T_n\le t_p$ index the
first and last FT searches at which the carcass is present, and let
$m_*:=\max\{n\ge m_0:~D_n=1\}$ index the last successful search (or
$m_*= m_0$ if it is never discovered).  Introduce the short-hand notation
$p_n(a,b):=\exp\big(-a- b(T_n-t_0)\big)$ for the probability of discovery
at the $n$th search, for $m_0\le n\le m^*$.  For a carcass that arrived in
an earlier search period to be discovered now it must have been
undiscovered and also ``bled through'' at each previous search.  Set
$D_n=1$ for a successful discovery and $D_n=0$ for a failure.  Then the
probability of the observed sequence of successes and failures for the
$k$th carcass, as a function of $(a,b,\bleed)$, is the sum over all
possible indices $m$ of the last search time $T_m$ at which the carcass
bleeds through,

\begin{subequations}\label{e:LH-discovery}
\begin{align}\begin{split}\label{e:LH-dis}
\cL^k_\mathrm{disc}(a,b,\bleed)
   = (1-\bleed) \sum_{m_*\le m<m^*} &\bleed^{m-m_0}
      \prod_{m_0\le n\le m} p_n(a,b)^{D_n}(1-p_n(a,b))^{1-D_n}\\
    + &\bleed^{m^*-m_0}\prod_{m_0\le n\le m^*} p_n(a,b)^{D_n}(1-p_n(a,b))^{1-D_n}.
\end{split}
\end{align}
The negative log likelihood contribution for all carcass combined is the
sum

\begin{align}\label{e:nllh-dis}
\ell_\mathrm{disc}(a,b,\bleed) &= \sum_k -\log\cL^k_\mathrm{disc}(a,b,\bleed).
\end{align}
\end{subequations}
The MLEs presented in \Sec{ss:srch} are the minimizing values
$(\hat a,\hat b, \hat\bleed)$.

\subsection{Posterior Distribution of Mortality}\label{ss:post-mort}
In this section we consider the posterior distribution of the mortality
$M_i$ in a fixed period $(T\imo,T_i]$ of length $I_i=(T_i-T\imo)$ days,
conditional upon the observed count $C_i$ in the search at time $T_i$, in
order to find interval estimates for $M_i$.  To make the notation less
cumbersome we omit the subscripts ``$i$''.

The total number $C$ of carcasses discovered in the search will in general
be a sum $C= C\on + C\oo$ of ``new'' carcasses that arrived during the
current interval and ``old'' ones that arrived in earlier periods, but were
undiscovered and unremoved in earlier periods.  In this model the mortality
$M\sim\Po(mI)$ in a particular search interval has a Poisson distribution
with uncertain mean $mI$ for a daily average rate $m\ge0$ which varies
sufficiently slowly from one interval to another that we may treat it as
constant over the arrival times of all the carcasses discovered in a
particular search.  We employ a Gamma prior distribution
$m\sim\Ga(\malp,\mbet)$ for $m$, usually with the Objective Bayes prior
parameters $\malp=1/2$, $\mbet=0$ \citep {Berg:Bern:Sun:2009}.

Each of the $M$ carcasses that arrive during the period has probability
$\Tst_0$ of being discovered in the current search, probability $(\oz)$ of
being discovered in some future search, and probability $(1-\Rst)$ of never
being discovered.  Thus the model may be described:

\begin{alignat*}2
    m    &\sim \Ga(\malp,~\mbet)&
         &\text{Average daily mortality}\\
   C\oo  &\sim \Po\big(m(\Rst-\Tst_0)I\big)&
         &\text{From all previous periods}\\
    M    &\sim \Po(mI)&
         &\text{Mortality this period}\\
   C\on \mid M &\sim \Bi(M, \Tst_0)&
         &\text{New count, conditional on $M$}\\
   C\on  &\sim \Po(m \Tst_0I)&
         &\text{New count, marginal}\\ 
   C     &= C\on + C\oo&\quad&\text{(new + old), indep.}
\end{alignat*}
where $\Rst\ge\Tst_0$ are given in \Eqns {e:Rst2}{e:Tst0}.

\subsubsection{Possible bleed-through ($\bleed>0$)}\label{sss:B>0}
First consider the case where $\Rst>\Tst_0$, and in particular $\bleed>0$,
so the mortality $M$ may take any nonnegative integer value--- even $M<C$,
since some or even all of the $C$ discovered carcasses may have arrived in
earlier search intervals.  Summing over the possible number $x=C\on$ of new
carcasses and integrating over the uncertain mean daily mortality $m$,

\begin{subequations}\label{e:CM}
\begin{align}
\P[ C&=C, M=M]\notag\\
 &= \sum_{x=0}^{C\wedge M} \int_0^\infty
    \set{\frac{m^{\malp-1}\mbet^\malp}{\Gamma(\malp)}e^{-m\mbet}}
    \set{\frac{(mI)^M}{M!}e^{-mI}}
    \set{\binom{M}{x}(\Tst_0)^x(1-\Tst_0)^{M-x}}\notag\\ &\qquad\times
    \set{\frac{[(\oz)mI]^{C-x}}{(C-x)!}e^{-{(\oz)mI}}}\,dm\notag\\
 &= \sum_{x=0}^{C\wedge M}
    \frac{\Gamma(\malp+C+M-x)}{\Gamma(\malp)(C-x)!(M-x)!x!}\quad
    \frac{\mbet^\malp (\Tst_0)^x(\oz)^{C-x}(1-\Tst_0)^{M-x} I^{C+M-x}}
         {[\mbet+(\oz+1)I]^{\malp+C+M-x} }\notag\\
 &= c \times \sum_{x=0}^{C\wedge M}
    \frac{\Gamma(\malp+C+M-x)}{(C-x)!(M-x)!x!}~z^x\notag\\
 &= c \times \frac{\Gamma(\malp+C+M)}
                  {C!\,M!}~{}_2F_1(-C,-M;1-\malp-C-M;-z)\label{e:C,M}
\end{align}
where ${}_2F_1(a,b;c;z)$ is Gauss' hypergeometric function
\citep[\S15]{DLMF:2014} and where

\[
 c = \frac{ \mbet^\malp (\oz)^C(1{-}\Tst_0)^M I^{C+M}}
         {\Gamma(\malp)[\mbet+(\oz+1)I]^{\malp+C+M}}\qquad
 z = \frac{\Tst_0[\mbet+(\oz+1)I]}{(1{-}\Tst_0)(\oz)\,I}.
\]
The induced marginal distribution of $C\sim\Po\big(m\Rst I)$ is negative
binomial,
\begin{equation}\label{e:C}
\P[ C=C] = \frac{\Gamma(\malp+C)}{\Gamma(\malp)~C!}
            \mbet^\malp(\Rst I)^C(\mbet+\Rst I)^{-\malp-C}.
\end{equation}\end{subequations}
Dividing \eqref{e:C,M} by \eqref{e:C} gives the conditional distribution
for mortality $M$ given a carcass count of $C$:

\begin{align}
  \P[M&=M\mid C=C] = c \times {}_2F_1(-C,-M;1-\malp-C-M;-z)\tag{\ref{e:M|C}}
\end{align}
with $c$ and $z$ given by

\begin{subequations}\label{e:M+C}\begin{align}
 \begin{split}\label{e:M|C-emp}
  c &= \frac
     {\Gamma(\malp+C+M)(\mbet+\Rst I)^{\malp+C}(\oz)^C(1-\Tst_0)^M\,I^M}
     {\Gamma(\malp+C)~M!~(\Rst)^C[\mbet+(\oz+1)I]^{\malp+C+M}}\\
  z &= \frac{\Tst_0~[\mbet+(\oz+1)I]} {(1{-}\Tst_0)(\oz)~I}.
\end{split}
\end{align}
For the Objective Bayes reference values $\malp=\half$ and $\mbet=0$ the
distribution is again given by \eqref{e:M|C}, but $c$ and $z$ are a bit
simpler and don't depend on the search interval length $I$:

\begin{align}
\begin{split}\label{e:M|C-obj}
  c &= \frac
     {\Gamma(\half+C+M)(\Rst)^{\half}(\oz)^C(1-\Tst_0)^M}
     {\Gamma(\half+C)~M!~(\oz+1)^{\half+C+M}}\\
  z &= \frac{\Tst_0(\oz+1)}{(1{-}\Tst_0)(\oz)}.
\end{split}
\end{align}\end{subequations}
This is well-defined even though, by \eqref{e:Ci-marg}, the marginal
predictive distribution of $C$ is degenerate for $\mbet=0$.  Objective
Bayes $100\gamma\%$ credible intervals for $M$ are presented in
\Sec{sss:OB} and illustrated in \Fig{f:post}, based on the conditional
distribution of $M$ (for specified $C$) given in \Eqns {e:M|C} {e:M|C-obj}.


\subsubsection{No bleed-through ($\bleed=0$)}\label{sss:B=0}
For the remaining case of $\Rst=\Tst_0$ where only ``new'' carcasses can be
found, the sum of \Eqn{e:C,M} reduces to the single term $x=C\le M$, so
only values $M\ge C$ are possible and for these \Eqn{e:M|C} becomes:

\begin{align}
 \P[M=M\mid C=C] = \frac{\Gamma(\malp+M)}{\Gamma(\malp+C)~(M-C)!}
(\Rst+\mbet/I)^{\malp+C}(1-\Rst)^{M-C}(1+\mbet/I)^{-\malp-M}. 
\tag{\ref {e:M|C-B=0}}
\end{align}
The number $M-C$ of undiscovered carcasses will have a negative binomial
conditional distribution
$(M-C)\mid C\sim\NB\big(\malp+C, (\mbet+\Rst I)/(\mbet +I)\big)$ or, for
the Objective Bayes case of $\malp=\half$, $\mbet=0$,
$(M-C)\mid C\sim\NB\big(C+\half, \Rst\big)$, justifying the interval
estimate for $M$ given in \Eqn {e:B0half}.

\need1\bibliography{statjour-abbr,acme}

\begin{thebibliography}{30}
\providecommand{\natexlab}[1]{#1}
\providecommand{\url}[1]{\texttt{#1}}
\expandafter\ifx\csname urlstyle\endcsname\relax
  \providecommand{\doi}[1]{doi: #1}\else
  \providecommand{\doi}{doi: \begingroup \urlstyle{rm}\Url}\fi

\bibitem[Berger et~al.(2009)Berger, Bernardo, and Sun]{Berg:Bern:Sun:2009}
J.~O. Berger, J.~M. Bernardo, and D.~Sun.
\newblock The formal definition of reference priors.
\newblock \emph{Ann. Stat.}, 37\penalty0 (2):\penalty0 905--938, 2009.

\bibitem[Bispo et~al.(2013{\natexlab{a}})Bispo, Bernardino, Marques, and
  Pestana]{Bisp:Bern:etal:2013a}
R.~Bispo, J.~Bernardino, T.~A. Marques, and D.~Pestana.
\newblock Discrimination between parametric survival models for removal times
  of bird carcasses in scavenger removal trials at wind turbines sites.
\newblock In J.~Lita~da Silva, F.~Caeiro, I.~Nat{\'{a}}io, C.~A. Braumann,
  M.~L. Esqu{\'\i}vel, and J.~Mexia, editors, \emph{Advances in Regression,
  Survival Analysis, Extreme Values, Markov Processes and Other Statistical
  Applications}, Studies in Theoretical and Applied Statistics, chapter 4, Part
  II. Springer-Verlag, 2013{\natexlab{a}}.
\newblock ISBN 978-3-642-34903-4.

\bibitem[Bispo et~al.(2013{\natexlab{b}})Bispo, Bernardino, Marques, and
  Pestana]{Bisp:Bern:etal:2013b}
R.~Bispo, J.~Bernardino, T.~A. Marques, and D.~Pestana.
\newblock Modeling carcass removal time for avian mortality assessment in wind
  farms using survival analysis.
\newblock \emph{Environmental and Ecological Statistics}, 20\penalty0
  (1):\penalty0 147--165, 2013{\natexlab{b}}.
\newblock \doi{10.1007/s10651-012-0212-5}.

\bibitem[Casella(1985)]{Case:1985}
G.~Casella.
\newblock An introduction to empirical bayes data analysis.
\newblock \emph{American Statistician}, 39\penalty0 (2):\penalty0 83--87, 1985.
\newblock \doi{10.2307/2682801}.

\bibitem[Erickson et~al.(1998)Erickson, Strickland, Johnson, and
  Kern]{Eric:Stric:etal:1998}
W.~P. Erickson, M.~D. Strickland, G.~D. Johnson, and J.~W. Kern.
\newblock Examples of statistical methods to assess risks of impacts to birds
  from wind plants.
\newblock In \emph{PNAWPPM-III: Proceedings of the Avian-Wind Power Planning
  Meeting {III}, San Diego, CA}. National Wind Coordinating Committee Meeting,
  May 1998, Washington, DC, 1998.
\newblock Prepared for the Avian Subcommittee of the National Wind Coordinating
  Committee by LGL, Ltd., King City, Ont.

\bibitem[Erickson et~al.(2001)Erickson, Johnson, Strickland, Young, Sernka, and
  Good]{Eric:John:Etal:2001}
W.~P. Erickson, G.~D. Johnson, M.~D. Strickland, D.~P. Young, Jr., K.~J.
  Sernka, and R.~E. Good.
\newblock \emph{Avian collisions with wind turbines: a summary of existing
  studies and comparisons to other sources of avian collision mortality in the
  {U}nited {S}tates}, 2001.
\newblock On-line at \url
  {http://www.west-inc.com/reports/avian\_collisions.pdf}.

\bibitem[Erickson et~al.(2005)Erickson, Johnson, and
  Young]{Eric:John:Youn:Etal:2005}
W.~P. Erickson, G.~D. Johnson, and D.~P. Young, Jr.
\newblock A summary and comparison of bird mortality from anthropogenic causes
  with an emphasis on collisions.
\newblock Technical Report PSW-GTR-191, U.S. Department of Agriculture,
  Washington, D.C., 2005.

\bibitem[Erickson et~al.(2008)Erickson, Jeffrey, and Poulton]{PSE:2008}
W.~P. Erickson, J.~D. Jeffrey, and V.~K. Poulton.
\newblock Puget sound energy wild horse wind facility post-construction avian
  and bat monitoring: First annual report january--december 2007.
\newblock Technical report, Puget Sound Energy (Ellensburg, WA 98296) and Wild
  Horse Wind Facility Technical Advisory Committee (Kittitas County, WA);
  prepared by Western EcoSystems Technology Inc, Cheyenne, WY, Jan 2008.

\bibitem[Etterson(2013)]{Ette:2013}
M.~A. Etterson.
\newblock Hidden {M}arkov models for estimating animal mortality from
  anthropogenic hazards.
\newblock \emph{Ecological Applications}, 23\penalty0 (8):\penalty0 1915--1925,
  2013.
\newblock \doi{10.1890/12-1166.1}.

\bibitem[Gelman et~al.(2009)Gelman, Carlin, Stern, and
  Rubin]{Gelm:Carl:Ster:Rubi:2009}
A.~Gelman, J.~B. Carlin, H.~S. Stern, and D.~B. Rubin.
\newblock \emph{Bayesian Data Analysis}.
\newblock Taylor \& Francis, Boca Raton, FL, 2nd edition, 2009.
\newblock ISBN 1-58488-388-X.

\bibitem[Howell and DiDonato(1991)]{Howe:DiDo:1991}
J.~A. Howell and J.~E. DiDonato.
\newblock \emph{Assessment of avian use and mortality related to wind turbine
  operations, {A}ltamont {P}ass, {A}lameda and {C}ontra {C}osta Counties,
  {C}alifornia, September 1988 through August 1989}, 1991.
\newblock Final Report to U.S. WindPower, Inc., Livermore, Calif.

\bibitem[Huso(2011)]{Huso:2011}
M.~M.~P. Huso.
\newblock An estimator of wildlife fatality from observed carcasses.
\newblock \emph{Environmetrics}, 22\penalty0 (3):\penalty0 318--329, 2011.
\newblock \doi{10.1002/env.1052}.

\bibitem[Huso et~al.(2014)Huso, Dalthorp, Dail, and
  Madsen]{Huso:Dalt:etal:2014}
M.~M.~P. Huso, D.~H. Dalthorp, D.~A. Dail, and L.~J. Madsen.
\newblock Estimating wind-turbine caused bird and bat fatality when zero
  carcasses are observed.
\newblock \emph{Ecological Applications}, pages~{\rm On--line only}, 2014.
\newblock \doi{10.1890/14-0764.1}.

\bibitem[Johnson et~al.(2003)Johnson, Erickson, Strickland, Shepherd, Shepherd,
  and Sarappo]{John:Eric:etal:2003}
G.~D. Johnson, W.~P. Erickson, M.~D. Strickland, M.~F. Shepherd, D.~A.
  Shepherd, and S.~A. Sarappo.
\newblock Mortality of bats at a large-scale wind power development at
  {B}uffalo {R}idge, {M}innesota.
\newblock \emph{The American Midland Naturalist}, 150\penalty0 (2):\penalty0
  332--342, 2003.
\newblock \doi{10.1674/0003-0031(2003)150[0332:MOBAAL]2.0.CO;2}.

\bibitem[Korner-Nievergelt et~al.(2015)Korner-Nievergelt, Behr, Brinkmann,
  Etterson, Huso, Dalthorp, Korner-Nievergelt, Roth, and
  Niermann]{KoNi:Behr:etal:2015}
F.~Korner-Nievergelt, O.~Behr, R.~Brinkmann, M.~A. Etterson, M.~M.~P. Huso,
  D.~H. Dalthorp, P.~Korner-Nievergelt, T.~Roth, and I.~Niermann.
\newblock Mortality estimation from carcass searches using the
  \texttt{R}-package \texttt{carcass} --- a tutorial.
\newblock \emph{Wildlife Biology}, 21\penalty0 (1):\penalty0 30--43, 2015.
\newblock \doi{10.2981/wlb.00094}.

\bibitem[Loss et~al.(2013)Loss, Will, and Marra]{Loss:Will:Marr:2013}
S.~R. Loss, T.~Will, and P.~P. Marra.
\newblock Estimates of bird collision mortality at wind facilities in the
  contiguous {U}nited {S}tates.
\newblock \emph{Biological Conservation}, 168:\penalty0 201--209, 2013.
\newblock \doi{10.1016/j.biocon.2013.10.007}.

\bibitem[Manville(2009)]{Manv:2009}
A.~Manville, II.
\newblock Towers, turbines, power lines, and buildings--- steps being taken by
  the {U.S. F}ish and {W}ildlife {S}ervice to avoid or minimize take of
  migratory birds at these structures.
\newblock In \emph{Proceedings of the Fourth International Partners in Flight
  Conference: Tundra to Tropics}, pages 262--272, 2009.

\bibitem[Olver et~al.(2010)Olver, Lozier, Boisvert, and Clark]{OLBC:2010}
F.~W.~J. Olver, D.~W. Lozier, R.~F. Boisvert, and C.~W. Clark, editors.
\newblock \emph{{NIST} Handbook of Mathematical Functions}.
\newblock Cambridge Univ. Press, New York, NY, 1.0.9 edition, 2010.
\newblock ISBN 978-0-521-19225-5.
\newblock URL \url{http://dlmf.nist.gov/}.
\newblock Print companion to \citep{DLMF:2014}.

\bibitem[P{\'e}ron et~al.(2013)P{\'e}ron, Hines, Nichols, Kendall, Peters, and
  Mizrahi]{Pero:Hine:Nich:etal:2013}
G.~P{\'e}ron, J.~E. Hines, J.~D. Nichols, W.~L. Kendall, K.~A. Peters, and
  D.~S. Mizrahi.
\newblock Estimation of bird and bat mortality at wind-power farms with
  superpopulation models.
\newblock \emph{Journal of Applied Ecology}, 50:\penalty0 902--911, 2013.
\newblock \doi{10.1111/1365-2664.12100}.

\bibitem[Pollock(2007)]{Poll:2007}
K.~H. Pollock.
\newblock Recommended formulas for adjusting fatality rates.
\newblock In \emph{California Guidelines for Reducing Impacts to Birds and Bats
  from Wind Energy Development}, pages 117--118, Appendix F. California Energy
  Commission, Renewables Committee, and Energy Facilities Siting Division, and
  California Department of Fish and Game, Resources Management and Policy
  Division, 2007.
\newblock {CEC} Final Report, Document CEC-700-2007-008-CMF.

\bibitem[Pyle and DeSante(2014)]{Pyle:DeSa:2014}
P.~Pyle and D.~F. DeSante.
\newblock \emph{List of {N}orth {A}merican birds and alpha codes according to
  {A}merican {O}rnithologists' {U}nion taxonomy through the 54th {AOU
  S}upplement [Updated 2014-09-29]}.
\newblock American Ornithologists' Union, Point Reyes Station, CA, 2014.
\newblock Available from \url{http://www.birdpop.org/alphacodes.htm}.

\bibitem[{R Core Team}(2015)]{R:2014}
{R Core Team}.
\newblock \emph{\texttt{R}: A Language and Environment for Statistical
  Computing}.
\newblock R Foundation for Statistical Computing, Vienna, AT, 2015.
\newblock URL \url{http://www.R-project.org}.

\bibitem[NIST DLMF()]{DLMF:2014}
NIST DLMF.
\newblock \emph{{NIST} Digital Library of Mathematical Functions}.
\newblock Release 1.0.9, 2014.
\newblock URL \url{http://dlmf.nist.gov/}.
\newblock Online companion to \citep{OLBC:2010}, at URL
  \url{http://dlmf.nist.gov/}.

\bibitem[Robbins(1955)]{Robb:1955}
H.~Robbins.
\newblock An empirical {B}ayes approach to statistics.
\newblock In J.~Neyman, editor, \emph{Proc.\ Third Berkeley Symp. Math.
  Statist. Prob.}, volume~1, pages 157--164. University of California Press,
  Berkeley, CA, 1955.
\newblock URL \url{http://projecteuclid.org/euclid.bsmsp/1200501653}.

\bibitem[Shoenfeld(2004)]{Shoe:2004}
P.~S. Shoenfeld.
\newblock Suggestions regarding avian mortality extrapolation.
\newblock On-line at \texttt{http://www.wvhighlands.org\slash Birds\slash
  Suggestions\-Regarding\-Avian\-Mortality\-Extra\-pola\-tion.pdf}, 2004.

\bibitem[Smallwood(2013)]{Smal:2013a}
K.~S. Smallwood.
\newblock Comparing bird and bat fatality-rate estimates among {N}orth
  {A}merican wind-energy projects.
\newblock \emph{Wildlife Society Bulletin}, 37\penalty0 (1):\penalty0 19--33,
  2013.
\newblock \doi{10.1002/wsb.260}.

\bibitem[Smallwood and Thelander(2005)]{Smal:Thel:2005}
K.~S. Smallwood and C.~G. Thelander.
\newblock \emph{Bird Mortality at the {A}ltamont Pass Wind Resource Area: March
  1998 -- September 2001}.
\newblock National Renewable Energy Laboratory (NREL), Golden, Colorado 80401,
  2005.
\newblock URL \url{http://www.osti.gov/bridge}.
\newblock Subcontract Report NREL/SR-500-36973.

\bibitem[Sovacool(2012)]{Sova:2012}
B.~K. Sovacool.
\newblock The avian benefits of wind energy: a 2009 update.
\newblock \emph{Renewable Energy}, 49:\penalty0 19--24, 2012.
\newblock \doi{10.1016/j.renene.2012.01.074}.

\bibitem[Warren-Hicks et~al.(2012)Warren-Hicks, Newman, Wolpert, Karas, and
  Tran]{WaHi:Newm:Etal:2012}
W.~Warren-Hicks, J.~Newman, R.~L. Wolpert, B.~Karas, and L.~Tran.
\newblock \emph{Improving Methods for Estimating Fatality of Birds and Bats at
  Wind Energy Facilities}, 2012.
\newblock URL
  \url{http://www.energy.ca.gov/2012publications/CEC-500-2012-086/CEC-500-2012-086.pdf}.
\newblock {C}alifornia Wind Energy Association publication CEC-500-2012-086.

\bibitem[Wilks(1938)]{Wilk:1938}
S.~S. Wilks.
\newblock The large-sample distribution of the likelihood ratio for testing
  composite hypotheses.
\newblock \emph{Ann. Math. Statist.}, 9\penalty0 (1):\penalty0 60--62, 1938.

\end{thebibliography}
\vfill\hfill{\tiny Last edited: \today, \now~EDT}
\end{document}